# The NIR Upgrade to the SALT Robert Stobie Spectrograph

Andrew I. Sheinis,[1,a] Marsha J. Wolf,[1,b] Matthew A. Bershady,[1] David A.H. Buckley,[2]
Kenneth H. Nordsieck,[1] Ted B. Williams[3]

[1] University of Wisconsin – Madison, Dept. of Astronomy, 475 N. Charter St., Madison, WI 53706
[2] South African Astronomical Observatory, Observatory 7935, South Africa
[3] Dept. of Physics and Astronomy, Rutgers University, Piscataway, NJ 08855

**ABSTRACT**

The near infrared (NIR) upgrade to the Robert Stobie Spectrograph (RSS) on the Southern African Large Telescope (SALT), RSS/NIR, extends the spectral coverage of all modes of the visible arm. The RSS/NIR is a low to medium resolution spectrograph with broadband imaging, spectropolarimetric, and Fabry-Perot imaging capabilities. The visible and NIR arms can be used simultaneously to extend spectral coverage from approximately 3200 Å to 1.6 μm. Both arms utilize high efficiency volume phase holographic gratings via articulating gratings and cameras. The NIR camera is designed around a 2048x2048 HAWAII-2RG detector housed in a cryogenic dewar. The Epps optical design of the camera consists of 6 spherical elements, providing sub-pixel rms image sizes of $7.5 \pm 1.0$ μm over all wavelengths and field angles. The exact long wavelength cutoff is yet to be determined in a detailed thermal analysis and will depend on the semi-warm instrument cooling scheme. Initial estimates place instrument limiting magnitudes at $J = 23.4$ and $H(1.4$-$1.6$ μm$) = 21.6$ for $S/N = 3$ in a 1 hour exposure.

## 1. INSTRUMENT CONCEPT

The Robert Stobie Spectrograph (RSS; formerly known as the Prime Focus Imaging Spectrograph, PFIS)[1,2] is currently in the commissioning phase on the Southern African Large Telescope (SALT).[3,4,5] It is the first facility spectrograph on the telescope. The RSS has a wide variety of capabilities including medium-band imaging, Fabry-Perot narrow-band imaging, long slit grating spectroscopy, multi-slit spectroscopy, and linear or circular polarimetry in any of the imaging or spectroscopic modes. RSS operates over a spectral band of 3200-9000 Å, with a future extension into the near infrared (NIR) envisioned from the beginning. The plan has always been to duplicate all capabilities of the visible arm, allowing simultaneous observations in the visible and NIR for all instrument modes.

The University of Wisconsin - Madison, together with University of North Carolina, Rutgers University and the South African Astronomical Observatory, proposes to build this NIR capability into the RSS. With this capability the RSS will be unique among instrumentation for 8-10 meter class telescopes in its ability to simultaneously record data in the visible and near infrared and it will open up a new window for the discovery and exploration of the most distant and earliest galaxies in the universe. The RSS/NIR upgrade will specialize in very high throughput, low to medium resolution spectroscopy, narrow-band imaging and spectropolarimetry starting at approximately 0.85 through 1.6-1.7 μm. The design includes an articulated camera, volume phase holographic (VPH) gratings and a double etalon Fabry-Perot system. This upgrade leverages the considerable effort and expense undertaken for the visible system, while preserving all of the visible capability.

The design of the RSS/NIR beam is an add-on to the visible system that has been envisioned since day one. The design philosophy uses most of the collimator for both the visible and NIR beams, with a final pair of achromats placed after a dichroic beamsplitter. Each achromat is optimized for a different band pass i.e visible and NIR. Each beam then feeds a separate disperser suite with articulating mechanism, camera and detector. Both beams will have an independent polarimetric capability and Fabry-Perot imaging capability.

---

[a] RSS/NIR Principal Investigator, sheinis@astro.wisc.edu
[b] RSS/NIR Project Scientist, mwolf@astro.wisc.edu

One of the unique design features of this instrument is that most of the NIR optical train will be cooled to only slightly below ambient temperature. Thermal emission due to the ambient temperature optics will significantly constrain the usable long-wave cutoff to be somewhere in the H band (1.6-1.7 µm). The design cutoff can be varied depending on bandpass and resolution. It is essentially determined by the requirement that the thermal emission of the optics (and telescope) be of order the intra-OH sky emission level.

## 2. OPERATIONAL MODES AND SCIENCE DRIVERS

### 2.1. Simultaneous Fabry-Perot NIR Imaging Spectroscopy and Visible Spectroscopy

The addition of the NIR double-etalon system on RSS NIR will place SALT at the forefront of astrophysics in the areas of cosmology, galaxy formation, and galactic and extra-galactic stellar and gas kinematics. SALT will be the only large telescope capable of visible and NIR Fabry-Perot imaging spectroscopy and will achieve resolutions from 500 to 12,500 over two full octaves in wavelength. This will provide a high-dispersion, high spatial resolution diffuse-object capability with a net efficiency that exceeds slit spectroscopy by more than an order of magnitude. Examples of new science made possible by this technology are:

- Detection of some of the earliest galaxies to form in the universe ($z=7-8$). By looking for Lyman alpha emission in narrow-band images at specific redshift ranges corresponding to the night-sky-line-quiet regions of the NIR spectrum, we will find and study young galaxies formed shortly after the Big Bang. Immediate (same-night, or same-run) follow-up with visible spectroscopic observations will allow for rejection of foreground objects. These observations will place SALT at the cutting edge of cosmology. A sample of the current narrow-band surveys typically find about 100 galaxies per square degree with Lyman-alpha line fluxes of a few times 1e-17 ergs/cm$^2$/s. The key goal is to try to measure sufficiently large samples so that one can measure the changes in the number density as a function of luminosity of the first galaxies and compare the results to predictions based on earlier structure in the cosmic microwave background. In addition, we will conduct targeted searches for forming galaxies near high-z QSO's in which we know the emission redshift.

- Stellar kinematics. With velocity precision of 1 km/s down to R=20, we have the ability to measure several thousand stellar velocities in about 2 hours of observing time. This enables studies of dynamical evolution, core collapse, and central black holes in Galactic globular clusters.

- Galaxy evolution. Fabry-Perot imaging spectroscopy allows the determination of internal dynamics of galaxies out to redshift of $z=3.3$, corresponding to a look-back time of 90% of the age of the universe. This will dramatically improve upon existing studies of the Tully-Fisher relation, cluster galaxy formation and evolution, and the role of hierarchical merging in galaxy formation.

- Sources of diffuse emission. With detection (S/N=3) of diffuse sources down to a surface brightness of 2 milli-Rayleighs in 15 minutes, RSS NIR enables studies of Galactic high velocity clouds, extragalactic HII regions, and Galactic ISM properties.

### 2.2. High Throughput, Medium Resolution NIR Spectroscopy

We will implement high throughput, medium resolution spectroscopy using VPH gratings. The extraordinarily high throughput of this instrument allows us to maximize the signal/noise gain from the SALT's large telescope aperture.

- The first stars. NIR spectroscopy of star forming regions and superstar clusters in the Small Magellanic Cloud will provide new data on low-metallicity star formation processes. This topic is very important for understanding the formation of the first stars in the universe.

- Star and planet formation. A remarkably hot topic in astronomy is the study of Brown dwarfs, young stars and circumstellar disks. Observations of the NIR excess in the spectra of brown dwarfs will allow us to determine whether this excess is due to circumstellar disks of planet-forming material.

- Brown dwarfs. One way to search for brown dwarfs is to look for the H-alpha emission which occurs in 30-40% of them due to accretion onto the stellar surface (Padaon 2005?), then immediate NIR spectroscopic follow-up. Most of these objects have been spectroscopically confirmed in the northern hemisphere, but many are left to be confirmed in the southern hemisphere, where SALT resides.

- Inner Galaxy. NIR spectral observations of the inner Galaxy will complement the GLIMPSE program (P.I., Ed Churchwell at Madison). In the NIR we will be able penetrate the dusty inner galaxy environment to study the ISM in HeI (1.0830 microns) and FeI (1.64micron).

- QSO absorption line studies of Lyman-α systems and Lyman break galaxies at much higher z and with better efficiency.

### 2.3. Spectropolarimetry

SALT will be unique in terms of spectropolarimetry, as there are no telescopes with apertures greater than 4 meters that have NIR spectropolarimetric capabilities. Only three other 8-10m telescopes have visible spectropolarimetric capability.

Simultaneous spectropolarimetry with the visible beam will for the first time allow wide wavelength coverage for rapidly variable polarimetric objects such as magnetic cataclysmic variables, pre-main sequence stars, novae, supernovae, and gamma ray bursts. The NIR spectropolarimetric imaging mode will allow penetration of the dusty inner Galaxy and thus magnetic field mapping of previously-obscured regions of the Milky Way. Fabry-Perot spectropolarimetry of scattered emission lines in dusty starburst galaxies will allow 3-D reconstruction of outflow from these galaxies.

- The capability for spectropolarimetry across the very large RSS Vis-NIR bandpass, 3200 Å to 1.6 μm (a factor of 5) will make it possible to separate interstellar from intrinsic polarization with very much increased reliability at SALT compared to other 10 m class telescopes. Interstellar polarization peaks in the visible, varying by only 30%, while it drops by a factor of 4 from the visible to 1.6 μm. Simultaneity in the visible and infrared will be particularly critical for variable objects.

- NIR spectropolarimetry will greatly increase the number of observable polarized spectropolarimetric targets (luminous stars; binary stars; pre-Main Sequence stars) in the Milky Way by allowing penetration towards the Galactic center.

- Embedded T-Tauri disks will be available to spectropolarimetry, allowing studies of much younger and more massive pre-Main Sequence stars.

- Spectropolarimetric studies of embedded AGN will be possible for the first time.

- Spectropolarimetry of supernovae at NIR wavelengths opens up study of asymmetries of ejecta of elements not easily seen in the visible, like Mg (0.92 and 1.09 microns) and FeII (1.57 microns).

## 3. OPTICAL DESIGN

### 3.1. Instrument Overview

An optical schematic of the RSS is shown in Figure 1. The visible and NIR arms of the RSS share the spectrograph slit, polarizing optics, and all elements of the collimator except the final one. A dichroic beam splitter sends the visible and NIR light to separate arms containing a final collimator optic optimized for each waveband. The next element in the NIR arm is a fold mirror that directs the light horizontally along the path to the disperser and the camera. A grating exchange mechanism selects the desired grating from a magazine and inserts it into a holder that rotates the grating to the correct articulation angle. The grating location is also the pivot point about which the entire camera articulates. Surrounding the grating area is a cold (-10 to -20 °C) baffle box to prevent stray light from reaching the disperser, to block light diffracted into unused orders, and to prevent warm surfaces from radiating to the grating. The grating holder, exchanger, and baffle box are all mounted on a kinematic plate that can be manually removed and replaced with the etalons and their insertion mechanism for the Fabry-Perot instrument mode. A pre-dewar that will be cooled down to -10 to -20 °C contains the first few optics of the camera. A cryo-cooler system will cool both the cold baffle box and the camera pre-dewar to slightly below ambient temperature. Finally, a cryogenic dewar contains the final camera optics, a short pass filter wheel, and the detector. The cryogenic dewar will have its own cooling system.

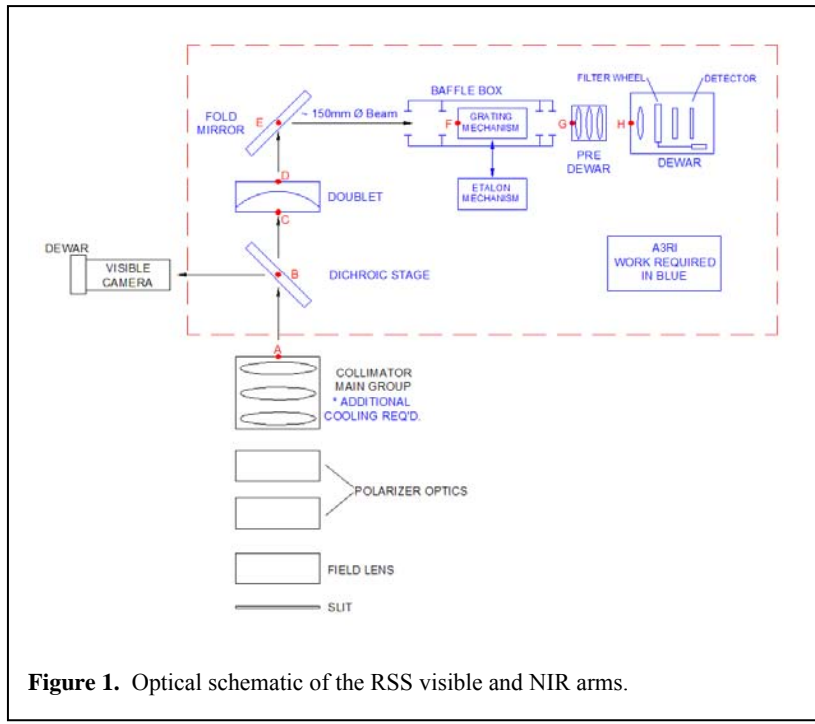

**Figure 1.** Optical schematic of the RSS visible and NIR arms.

### 3.2. Camera Optics

We have contracted the services of Dr. Harland Epps, the renowned optical designer from Lick Observatory to produce a preliminary optical design for the RSS NIR camera. Two all-spherical, 6-element optical designs are shown in Figure 2. The primary difference between the two is the use of a single $BaF_2$ element in one of the designs. The distribution of optical power among the lens elements is also shown in the figure. Parallel ray bundles of light radiate from the entrance pupil (not shown) 200 mm on the left, moving toward the right. Those which will form an image at full-field can be seen as they appear on the first lens surface. The rays pass through the lens elements and converge to focus at the flat detector array on the right. The total length from the first lens surface to the detector , shown as "Total Track" on the figure , is 484.85 mm for camera 1 and is 439.40 mm for camera 2, which are substantially more than the camera's 302.0-mm focal length.

When camera 1 is illuminated in perfectly parallel light as described above it produces residual aberrations with an rms image diameter of 7.5 +/- 1.0 μm averaged over all field angles and wavelengths within the 0.85 to 1.6 μm passband without refocus, with 4.7 μm of maximum rms lateral color (Figure 3), while camera 2 produces residual aberrations with an rms image diameter of 10.2 +/- 2.0 μm averaged overall field angles and wavelengths. Maximum 3rd-order barrel distortion for camera 1 is 0.29% at the edge of the full field (the corners of the detector) and 0.35% for camera 2. The differences not withstanding, it seems fair to say that both camera alternatives are extremely well corrected such that from a practical standpoint, they are virtually "perfect" with respect to the 18-micron pixel size in the intended

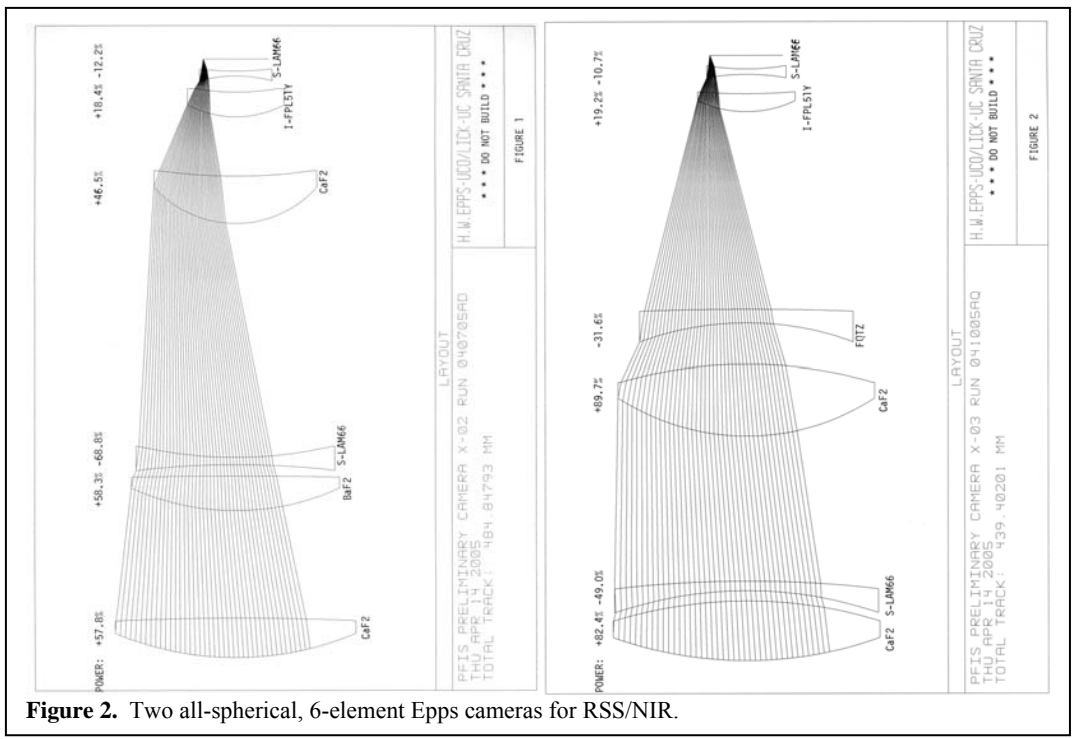

**Figure 2.** Two all-spherical, 6-element Epps cameras for RSS/NIR.

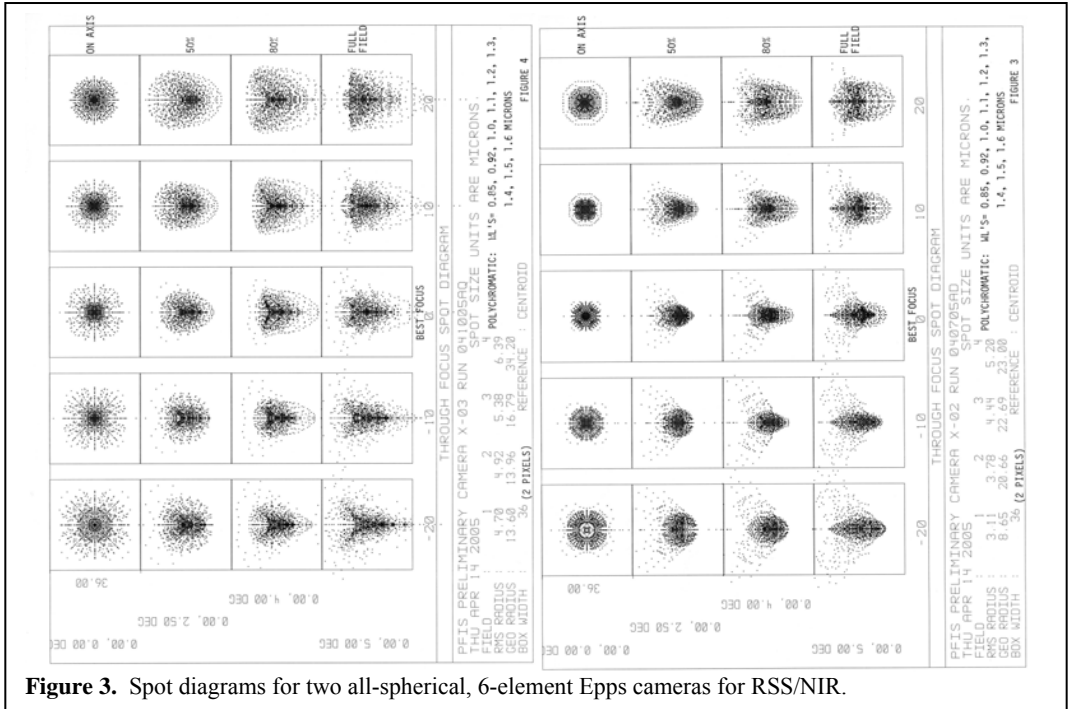

**Figure 3.** Spot diagrams for two all-spherical, 6-element Epps cameras for RSS/NIR.

HAWAII-2RG array.

### 3.3. Dispersive Elements

The proposed design takes advantage of several new technologies. VPH gratings, which are formed by exposing a gelatin material to a laser interference pattern, have a number of important advantages over conventional "surface relief" transmission and reflection gratings. Very high efficiency VPH transmission gratings are available, which allows a very compact, efficient, all transmissive system with a simpler camera. Visible VPH gratings are now in hand for RSS. The VPH gratings for RSS/NIR will be among the first developed for the NIR.

Rotating the grating and the camera "tunes" the VPH blaze efficiency peak over a factor of two in wavelength, providing the highest possible efficiency with a small number of gratings. (Some potential grating configurations are shown in Figure 4.) This does, however, entail the complication of an articulation mechanism for the camera. Because the gratings can be used efficiently at a very high angle, the spectrometer can produce a higher dispersion than a conventional reflection-grating spectrometer from a moderate beam diameter. Thus allowing the spectrometer to be as compact as possible.

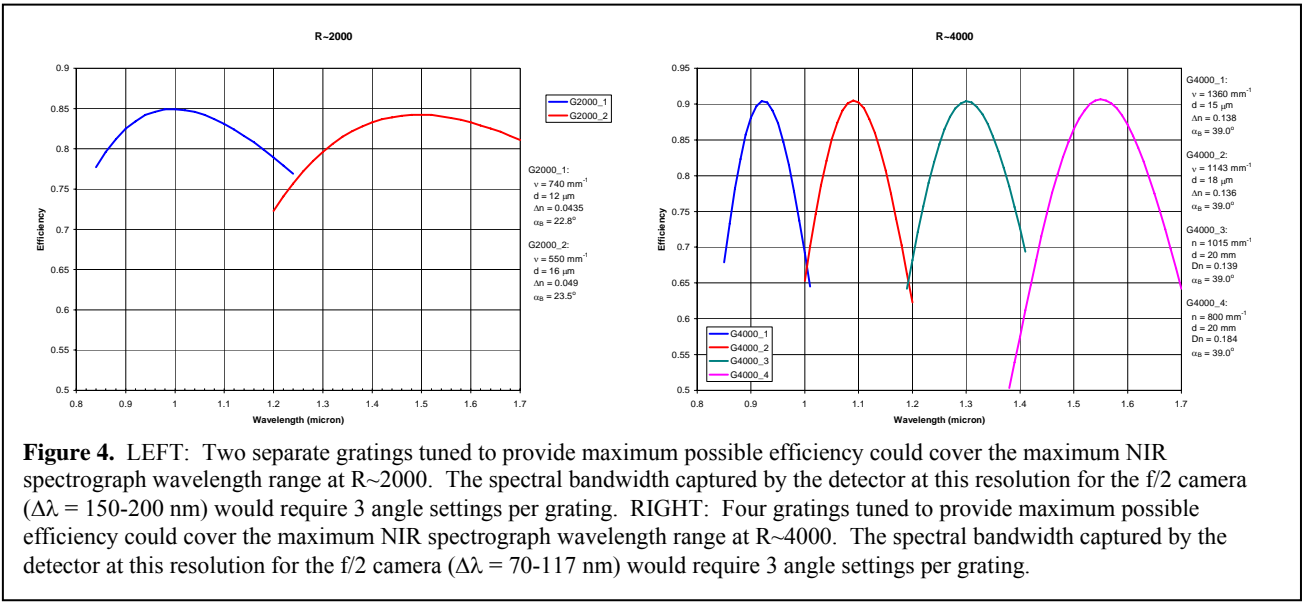

**Figure 4.** LEFT: Two separate gratings tuned to provide maximum possible efficiency could cover the maximum NIR spectrograph wavelength range at R~2000. The spectral bandwidth captured by the detector at this resolution for the f/2 camera (Δλ = 150-200 nm) would require 3 angle settings per grating. RIGHT: Four gratings tuned to provide maximum possible efficiency could cover the maximum NIR spectrograph wavelength range at R~4000. The spectral bandwidth captured by the detector at this resolution for the f/2 camera (Δλ = 70-117 nm) would require 3 angle settings per grating.

## 4. MECHANICAL COMPONENTS

Where possible, mechanical components from the RSS visible arm will be duplicated in the NIR arm. Such items include the filter exchange mechanism, etalon insertion mechanism, and polarizing beam splitter insertion mechanism. Major components are briefly described below.

### 4.1. Support Structure

The RSS space-frame structure was designed from the beginning to accommodate the NIR beam. This invar structure, which was designed and fabricated at Rutgers, contains all the structural mounting locations required for the NIR, articulating mechanism, camera radius slide, etalon insertion mechanism, grating magazine and exchange mechanism.

The RSS structure, shown in Figure 5, consists of an open-truss welded invar structure. It is designed to accommodate the following key components:

1) The RSS structure must be attached to the Prime Focus Instrument Platform, which is an 1800 mm diameter ring.

2) The collimator tube that holds all the collimator optics and is in the center of the instrument. The collimator is stationary.

3) Two cameras, the visible and NIR cameras. These cameras are at right angles to the collimator and also articulate around the VPH grating disperser. They each articulate to a maximum angle of 100°.

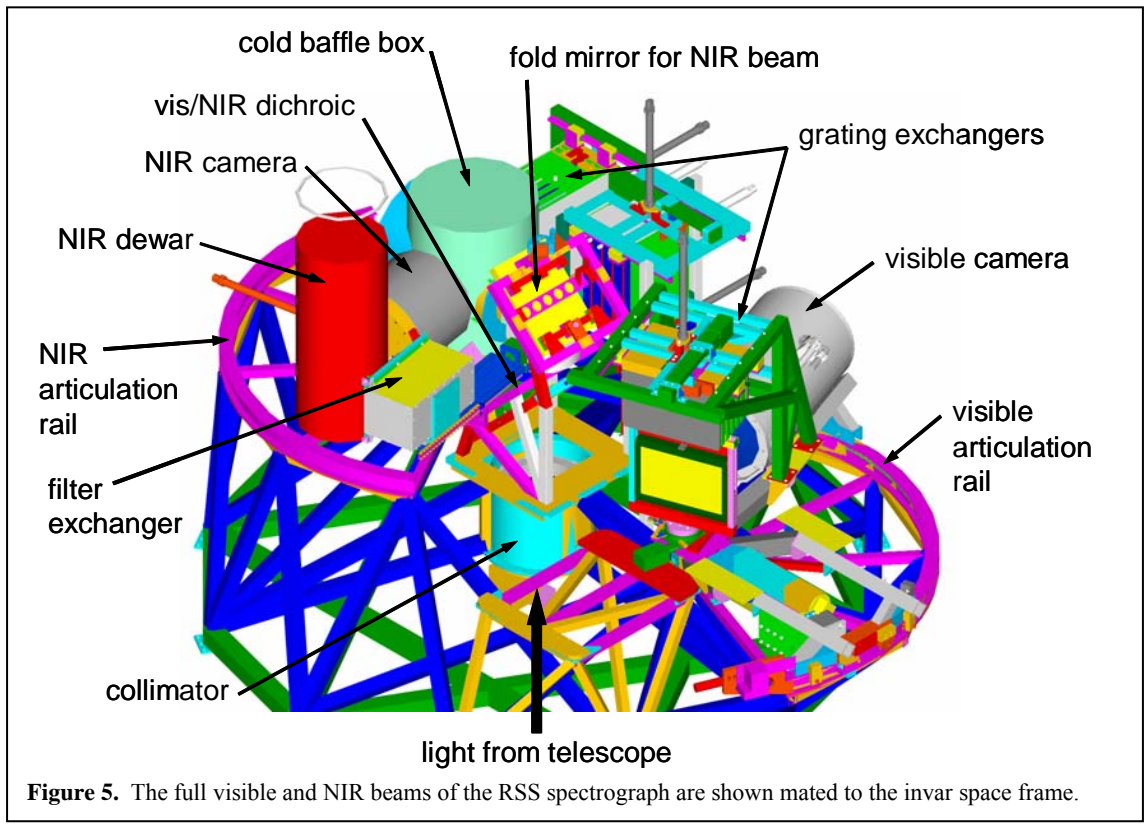

**Figure 5.** The full visible and NIR beams of the RSS spectrograph are shown mated to the invar space frame.

## 4.2. Grating Exchange Mechanism

The design concept for the grating exchange mechanism is shown in Figure 6. The grating mechanism inserts one of 4-6 gratings into the beam at the pupil. When one of the gratings is in position, the grating rotator rotates it to the required angle for the observation. The grating mechanism consists of a magazine on a frame and slide, an insertion pneumatic and the grating rotator stage. The result of the conceptual design study indicates that the visible grating change mechanism concept will be suitable for the NIR. It will require repackaging and modification to allow it to work in a horizontal configuration. Furthermore, due to space constraints, it will be repackaged onto a kinematic platform so that it may be manually removable for exchange with the NIR Fabry-Perot etalon and insertion mechanism on a similar platform. This configuration will require that grating spectroscopy or Fabry-Perot imaging be done in campaign modes on the telescope.

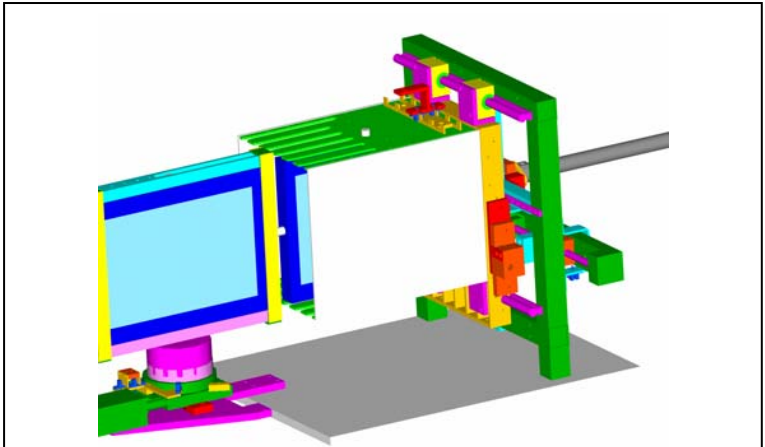

**Figure 7.** The design concept for the grating exchange mechanism. This mechanism is essentially a reconfiguration of the existing mechanism to work in a horizontal mode. It has the feature that is mounted to a kinematic platform, allowing for a manual exchange of this mechanism with the Fabry-Perot etalon and associated insertion mechanism.

## 4.3. Cold Baffle Box

A cold baffle-box which envelops the grating receiver will be developed to minimize the diffraction and scattering of the warm instrument-wall radiation into the beam. The packaging study has identified sufficient space available. This component will be actively cooled via a closed-cycle cryo-system that also cools the pre-dewar containing the first few camera optics. The concept for this cold baffle is shown in Figure 7.

## 4.4. Articulation Mechanism

The articulation mechanism (Figure 8) rotates the camera tube (and all that is attached to it – detector, filter mechanism, beamsplitter inserter) from 0 to 100 degrees. The articulation mechanism consists of the cradle and frame that supports the camera tube on the articulation bearing and I-beam rail, a motor/gearbox drive assembly and a precision positioning detent assembly. This mechanism will be copied in concept, but will be re-speced for the higher load requirements of the NIR camera and dewar.

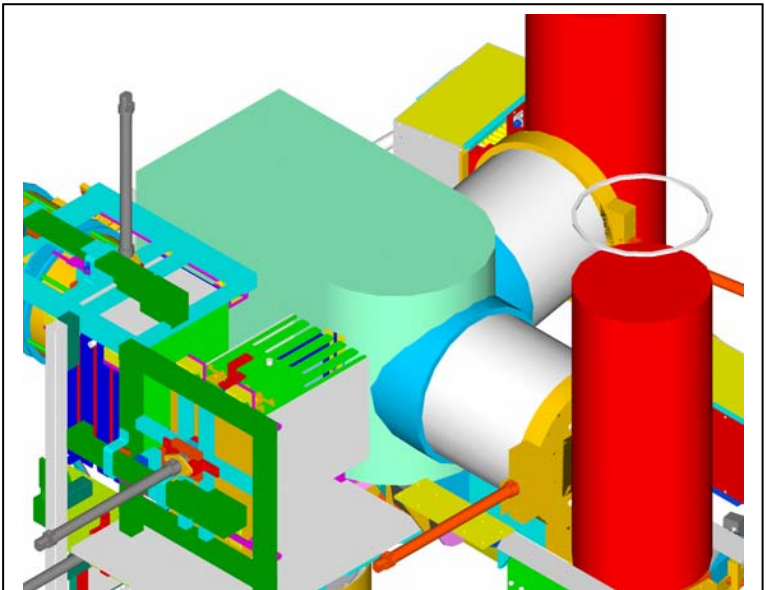

**Figure 6.** Cold baffle-box, shown with camera assembly at both articulation extremes and the grating exchange mechanism.

## 4.5. Etalon Assembly

The etalon assembly will be mounted to the kinematic platform to allow exchange with the grating mechanism. This will allow either of these two modes to be available in campaign mode at the telescope. The mechanism itself will be duplicated from the visible design. The RSS/NIR Fabry-Perot (FP) system will use 1 or 2 servo-controlled etalons. This decision will be based on the science requirements and cost trade-off.

Within the etalon, positioners set the parallelism and gap of the etalon plates, and the plate positions are monitored by capacitance sensing. This design provides high stability and repeatability. The spectral resolution of an etalon is set by the size of the spacing between its plates and by their reflectivity; this resolution is fixed for a given etalon, although the lowest resolution etalons have small enough gaps that they can be tuned by their piezos through approximately a factor of two in resolution.

Depending on the resolution and spectral coverage requirements, 15-30 interference filters will be required to isolate the FP orders over the entire spectral range. These will be installed in the below-described magazine with a capacity of 14 filters, which will cover the required observations for a given night.

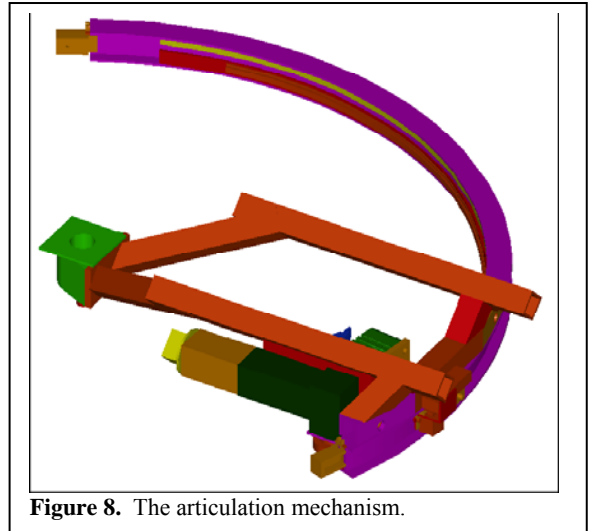

**Figure 8.** The articulation mechanism.

### 4.6. Camera

The RSS/NIR camera (Figure 9) consists of 2 main subsystems: 1) a cryogenic dewar, containing the detector (and associated focus mechanism if required), 2-3 of the camera lens elements, and a several (of order 5) position short-pass filter mechanism; 2) An atmospheric-pressure pre-dewar containing the remaining camera lenses and the interference-order-blocking filter, delivered by the filter exchange mechanism.

The evacuated cryogenic dewar will utilize closed cycle cryo-cooler technology to cool the detector to cryogenic temperature. It will either be purchased commercially or developed by a yet-to-be determined partner. The baseline design uses a modified stock dewar from IR Labs in Tucson, AZ.

The pre-dewar will be back-filled with dry gas at atmospheric pressure and use a separate cryo-cooler to chill its optics and dry gas to several 10's of degrees C below ambient. The order-blocking interference filters will be placed automatically within the pre-dewar and will be passively cooled by the dry gas.

### 4.7. Filter Exchange Mechanism

The filter mechanism is the mechanism that selects one of 14 filters from the filter magazine and inserts it into the beam. The visible filter mechanism will be duplicated from the visible arm with little or no modification. Filters will be passively cooled by placing them in the pre-dewar.

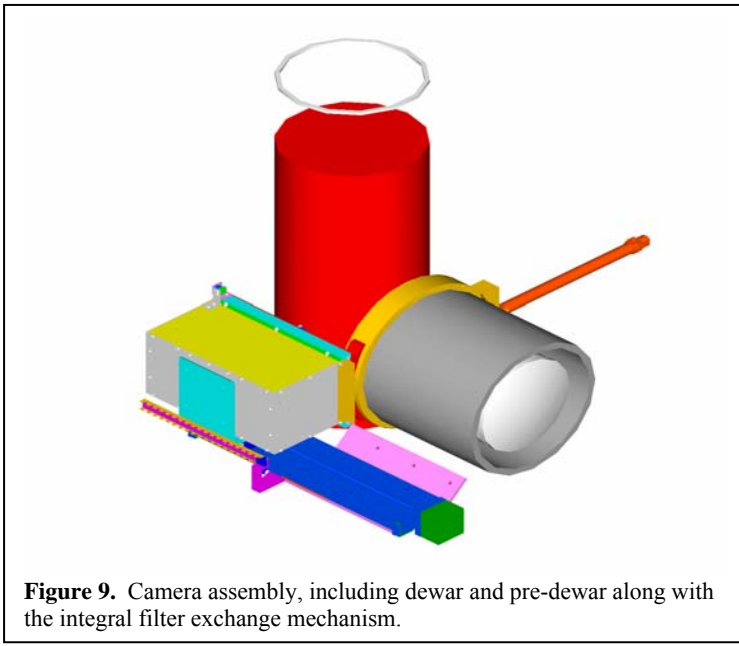

**Figure 9.** Camera assembly, including dewar and pre-dewar along with the integral filter exchange mechanism.

## 5. DETECTOR

This instrument will be based on the Rockwell HAWAII-2RG chip. This 4 megapixel HgCdTe focal plane array will be custom cold-filtered to the 0.85 to 1.7 µm region. Filtering combined with a custom AR coating on the focal plane will allow the spectral response of the chip to be tailored to maximize signal-to-noise. While the baseline approach will be to use a modified dewar and electronics package from IR Labs in Tucson AZ, it is desirable to evaluate detector/cryostat packages as in kind contribution from new partners.

Rockwell Scientific's new 2048 x 2048 HAWAII-2RG™ is the state of-the-art multiplexer for advanced astronomy and space telescope applications. Providing the ability to choose detector material (HgCdTe or Silicon PIN) allows the user access to any band from 350 nm to 5.3 μm. Selectable number of outputs (1, 4 or 32) and user selectable scan directions provide complete flexibility in data acquisition. The "Guide Mode" provides a programmable window which may be read out continuously at up to 5 MHz pixel rate for stable tracking of guide stars. The readout is designed to allow interleaved readout of the guide window and the full frame science data.

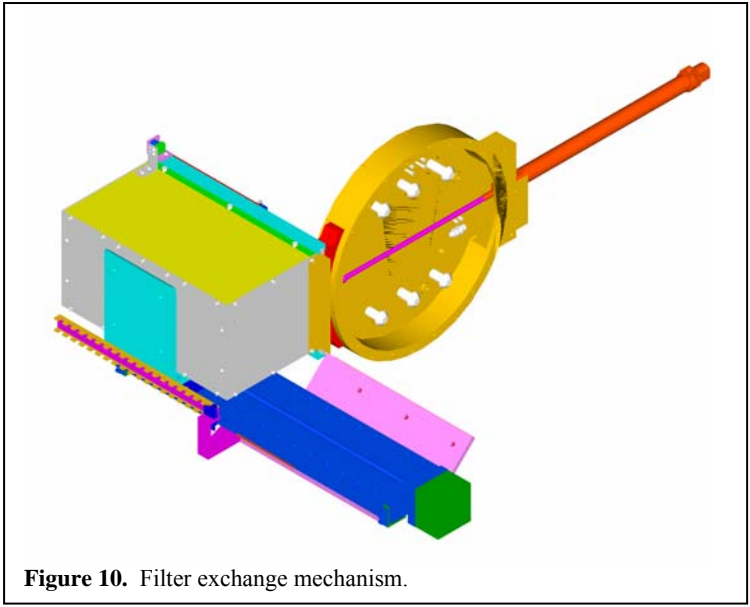

**Figure 10.** Filter exchange mechanism.

## 6. THERMAL ANALYSIS

### 6.1. Initial Estimates

The limiting magnitude expected for stellar observations in the NIR spectroscopic mode with the non-cryogenic spectrograph has been analyzed. At a fixed signal-to-noise ratio (SNR) requirement, the log wavelength cutoff will vary as a function of optics temperature and noise rejection. Cost and weight implications necessarily point the design to two possibilities: ambient temperature design, or a cooled, below-ambient design. A fully cryogenic instrument is not a viable option due to weight and budget considerations and most importantly that it would require removal of the visible instrument. Additional cost for the below-ambient instrument are not more than 10% of the instrument budget. The conclusion is that the RSS/NIR beam can be fully competitive and in many cases somewhat better than the best NIR imaging spectrographs in the J-band, achieving a limiting magnitude in 1 hour that is fully dependent on the sky brightness at our site. The instrument magnitude limit under these conditions is 23$^{rd}$ magnitude or better, approximately ¾ of a magnitude better than comparable systems on 8-10 meter class scopes. With careful attention to stray light issues and thermal emission from the optics, this spectrograph can be fully competitive in the bottom of H-band. The long wavelength cutoff through which the instrument remains sky-limited is dependent on the instrument details. For the reasonable instrument modeled (R=4000), we see that a cutoff at 1.6 μm gives a sky limited limiting magnitude of 20.5 and an instrumental limit of 21.6, which is competitive with the best instruments available. Moving the cutoff for the nominal instrument to 1.7 μm drops the instrumental limit 20.6 bringing the total limit down to 20.3.

Performance modeling has been done to estimate the red cutoff required to achieve a sky-noise-limited observation in 1 hour. This criterion is reached when the magnitude limit due to total instrument noise is 0.5 magnitudes fainter than the sky limit. The calculation is based on sky spectra for Mauna Kea provided by Gemini Observatory. The sky limit in the H-band occurs at approximately 20.4 to 20.6. Results are summarized in Table 1. The model is based on the following parameters.

- SNR requirement of 3, an integration time of 1 hour
- Ambient temperature of 0 °C, a pre-Dewar temperature of -20 °C
- Cryogenic temperature of 77 K
- Spectral resolution of R=4000
- Average grating efficiency 80% over entire band

**Table 1.** Limiting magnitudes at different long wavelength cutoffs and instrument cooling schemes.

|  | J | H cutoff 1.6 | H cutoff 1.65 | H cutoff 1.7 | H cutoff 1.7 |
|---|---|---|---|---|---|
| Ambient Optics temp (degreesC) | 0 | 0 | 0 | 0 | -10 |
| cold optics temperature (degreesC) | -20 | -20 | -20 | -20 | -20 |
| Instrumental Magnitude Limit | 23.4 | 21.6 | 21.1 | 20.6 | 21.1 |
| Sky magnitude limit | 20.3 | 20.5 | 20.5 | 20.6 | 20.6 |
| Combined magnitude limit | 20.3 | 20.5 | 20.6 | 20.3 | 20.6 |

Red cutoffs are listed below along with engineering changes required to achieve them, in order of increasing cost and complexity.

**1.54 μm:** The instrument becomes sky-limited with a red-cutoff of approximately 1.54 μm for the base-line instrument, i,e, no narcissus mirrors, no additional cooling beyond the dewar and pre-dewar housing the camera and detector.

**1.62 μm:** Adding a gold-coated slit, with effective emissivity of 0.1 and gold coating the moving baffle (the moving baffle masks portions of the telescope pupil that fall off the primary mirror during object tracking), blocking 50% of the warm observatory radiation increases the red coverage to 1.62 μm. Costs for this stage are very small compared to the instrument cost. This can achieved by gold coating the moving baffle and the bottom of the instrument payload as seen by the primary, with gold foil. Gold foil is available for about $100/sq ft. Costs for gold-coating the slit are based on discussions with Epner technologies in NY. They estimate a coating cost of $250 per slit mask for the resident slit-masks and $50-$100 /slit mask for visiting masks done in volume.

**1.66 μm:** Reducing the ambient instrument temperature to -10 °C further brings the red limit down to 1.66 um. This can be achieved by wrapping the collimator barrel with glycol cooling lines and insulating. Minimum achievable temperature will be determined by optical design constraints and opto-mechanical issues. Alan Schier of the Pilot group, who designed the collimator, suggests a retrofitted cooling system for the collimator would be in the $5K - $10K range. He believes that the internal stress-limit specification in the optics can be maintained down to a temperature of -20 °C.

**1.71 μm:** A red cut-off of 1.71 μm can be achieved by reducing the ambient temperature down to the opto-mechanical design limit of -20 °C, reducing the slit emissivity to 0.03 and introducing an internal cold-baffle at this same temperature to block ½ of the unwanted grating-coupled radiation. This would simply be a cold, gold-coated box surrounding the active area of the grating, with entrance and exit apertures for the 1$^{st}$ order radiation. Initial cost estimates are approximately $10K-$15K USD.

Based on these arguments, it is likely that the cost of forging deep into the H-band with a semi-warm spectrograph can be kept to less than an additional 5-10% of the total cost for a J-band system, exclusive of the differences in the raw detector costs. Nonetheless, these estimates were based on assumptions about the effectiveness of the stray light rejection. More accurate estimates will require a full stray light analysis of the system.

**Table 2.** Limiting magnitudes of existing or planned 8-10 meter class NIR spectrographs. Note: The first four instrument entries are based on modeling alone.

| name | Instrument | Limiting Magnitudes | Resolution l/dl | Fabry perot | Polarimetry |
|---|---|---|---|---|---|
| SALT | PFIS/NIR | J=23, H(1.4-1.6 um)=21.2, snr=3,r=4000,t=1hr | 1000-10,000 | yes | yes |
| Keck | Nirspec | J=22.2-21.5, SNR=10, t=1hour, R=2000 | 2000, 25,000 | | |
| GTC | EMIR | J=22.2,H=21,K=20.2 for SNR=5;t=1 hours | 5000(J), 4250(H), 4000(K) | | |
| HET | MRS beam 2 | | 5000, 20,000 | | |
| HET | LRIS-J | J~21-22, SNR=5, t=1 hr | | | |
| Subaru | OHS | | 300-1000 | | |
| | IRCS | J=16.1 mag/arcsec2 5 sigma detct at R=3790 | 200-20000 | | |
| | CIAO | | 300-1200 | | |
| ESO | CRIRES | J=17, R=20,000, SNR=3, 1hour | 20K-100K | | |
| ESO | ISAAC | J=19.5, SNR=5, 1hour, R=3000 | 500-3000 | | yes |
| ESO | SINFONI | | 2000-4000 | | |
| ESO | CONICA | | | yes | yes |
| Gemini N | NIRI | J=19.3, SNR=5, 1 hour | 460-1650 | | k-band |
| Gemini N | NIFS | | 5300 | | |
| Gemini S | Phoenix | | 50K-75K | | |
| Gemini S | GNIRS | J=19.7 (R=1700), J=18.4 (R=5900), SNR=5, 1hour | 1700-18000 | | yes |

Based on the above estimate, we expect that in the J-band the limiting magnitude of RSS/NIR may be significantly better than all other systems in the J-band and competitive in the bottom of the H-band (see Table 2). However, what is quoted for RSS/NIR in the J-Band is the limiting magnitude due to instrumental noise only. The relative success of RSS/NIR will depend entirely on the site. If the SALT site is darker then we may have the most sensitive spectrometer in the J-band. This is primarily due to throughput achieved using a VPH grating along with very high transmission optics.

### 6.2. Planned Thermal and Stray Light Analysis

Although the level of modeling described above, gives us an initial estimate of how far into the H-band we may be able to go, a more detailed study is necessary to steer the actual instrument design. Unexpected thermal background issues have plagued previous attempts to implement semi-warm NIR spectrographs. We plan to perform a detailed stray light and thermal analysis of RSS/NIR using the commercial software, ASAP by Breault Research Organization. The software is unique to other optical design programs in that it allows the importation of both optical and mechanical instrument designs. Surface features of optics, mounts, baffling, and all surrounding surfaces can be specified and temperatures assigned to each object, including thermal radiation. The software performs a non-sequential ray trace, allowing reflections and emission from every surface in all directions. Rays of light reaching the detector and generating "hot spots" can be traced back to their origin, identifying problem areas that can be modified in the mechanical design to minimize the ultimate thermal background of the instrument. Such an analysis of RSS/NIR should catch many of the problems that otherwise might lie in wait to surface during laboratory testing. With this analysis we should be able to reduce extra time required for empirical modifications in the lab by identifying and addressing problems up front during the design phase.

### 7. SUMMARY

With the addition of a NIR arm the RSS will be unique among instrumentation for 8-10 meter class telescopes in its ability to simultaneously record data in the visible and near infrared. The RSS/NIR upgrade will specialize in very high throughput, low to medium resolution spectroscopy, narrow-band imaging and spectropolarimetry starting at approximately 0.85 through 1.6-1.7 μm. High throughput will be achieved by using VPH gratings and high transmission optics.

A fully cryogenic NIR spectrograph was not an option because of the configuration in which the NIR arm shares parts with the visible spectrograph. Therefore, much of the NIR spectrograph is cooled only slightly below ambient temperature. Careful thermal analysis will be undertaken to guide the details of this cooling and baffling to provide sufficient rejection of thermal radiation that would otherwise severely limit instrument performance. Initial estimates place instrumental limiting magnitudes at J=23.4 and H(to 1.7 μm)=21.1 in the design with maximum cooling. The preliminary system specifications are summarized below.

| | |
|---|---|
| Telescope f/# | f/4.18 |
| Telescope aperture | 10 m |
| Collimator Focal length | 622.7 mm |
| Camera Focal length | 302 mm |
| Final f/# | 2.025 |
| Final platescale | 108 μm/arcsec |
| Final platescale | 6.0 pixels/arcsec |
| Detector | HAWAII-2RG |
| Pixels | 2048 X 2048 x 18 μm |
| Field of View | 8.0 arcmin diagonal |
| Nominal spectral range | 0.85 - 1.7 μm (dependent on R and optics temperature) |
| Possible modes of operation | 0.35-1.4 μm vis-NIR band or 0.4-1.6 μm vis-NIR plus H-band in a single exposure using both legs of the spectrometer. |
| Maximum dispersion | 11,783 with 0.75 arcsec slit at 110 degree grating angle (this is spectrometer limit, independent of grating choice) |
| Limiting Magnitudes | 20.3 in J-band and 20.6 in H-band |

# REFERENCES

1. H.A. Kobulnicky, K.H. Nordsieck, E.B. Burgh, M.P. Smith, J.W. Percival, T.B. Williams, D. O'Donoghue, "The Prime Focus Imaging Spectrograph for the Southern African Large Telescope: Operational Modes," *Proc. SPIE* 4841, 1634 (2003).
2. E.B. Burgh, K.H. Nordsieck, H.A. Kobulnicky, T.B. Williams, D. O'Donoghue, M.P. Smith, J.W. Percival, "The Prime Focus Imaging Spectrograph for the Southern African Large Telescope: Optical Design," *Proc. SPIE* 4841, 1463 (2003).
3. J.G. Meiring, D.A.H. Buckley, M.C. Lomberg, R.S. Stobie, "Southern African Large Telescope (SALT) Project: Progress and Status After 2 Years," *Proc. SPIE*, 4837, 11 (2003).
4. A. Swat, D. O'Donoghue, J. Swiegers, L. Nel, D.A.H. Buckley, "The Optical Design of the Southern African Large Telescope," *Proc. SPIE*, 4837, 564 (2003).
5. D.A.H. Buckley, J.B. Hearnshaw, K.H. Nordsieck, D. O'Donoghue, "Science Drivers and First Generation Instrumentation for the Southern African Large Telescope (SALT)," *Proc. SPIE*, 4834, 264 (2003).